\documentclass[11pt,a4paper]{article}
\usepackage[utf8]{inputenc}
\usepackage[T1]{fontenc}
\usepackage[margin=2.5cm]{geometry}
\usepackage{setspace}
\usepackage{graphicx}
\usepackage{booktabs}
\usepackage{tabularx}
\usepackage{array}
\newcolumntype{R}{>{\raggedleft\arraybackslash}X}
\usepackage{xcolor}
\usepackage{url}
\makeatletter\g@addto@macro{\UrlBreaks}{\UrlOrds}\makeatother   
\usepackage[natbibapa]{apacite}
\usepackage[colorlinks=true,allcolors=blue]{hyperref}
\pdftrailerid{}
\hypersetup{pdfcreator={},pdfproducer={},pdfauthor={},pdftitle={},pdfsubject={},pdfkeywords={}}
\providecommand{\doi}[1]{\url{https://doi.org/#1}}
\AtBeginDocument{}
\AtBeginDocument{}   

\begin{document}
\shortcites{rfc9309,longpre2024,liu2025,kim2025,hoffmann2026,dinzinger2024survey,steinacker2026,lopez2026,sun2007,sun2010,lepochat2019tranco,plantin2018,ulloa2025,kajava2025,ragnedda2022,urman2022,utz2019,nouwens2020,matte2020,mathur2019,hils2020,hils2021,scheitle2018,scheerder2017,nakagomi2026,dahlke2023,weiss2025,sen2021,daikeler2025,gaughan2026,holzl2026,respi2026}   
\sloppy

\begin{center}
{\Large\bfseries Who sets the consent signal? Platform defaults and crawl coverage in measuring opt-outs from AI training\par}
\vspace{1em}
{\large Marcin Lawenda\par}
\vspace{0.3em}
{\small Poznan Supercomputing and Networking Center, Poznań, Poland\\
\href{mailto:lawenda@man.poznan.pl}{lawenda@man.poznan.pl}\par}
\vspace{0.3em}
{\small Preprint, November 2026\par}
\end{center}

\begin{abstract}
\noindent Web data used in social research and in the training of AI models increasingly carry consent signals, but who sets these signals is rarely measured. This paper attributes opt-outs from AI training to the party that set them. The complete lists of Polish, Czech, Slovak, Hungarian and German domains recorded by Common Crawl, with 3,445,751 reachable sites, were crawled in October 2026 together with a random sample of 10,000 Slovak registry domains. Each signal was matched against vendor templates and identical files. Four exploratory hypotheses were examined with logistic models. Between 0.70\% and 3.82\% of the listed sites opt out. Platform and hosting defaults set 15\% to 78\% of the opt-outs. Settings active by default reach 60\% to 95\% of a vendor's sites, while settings that owners switch on reach at most 2.1\%. Most of the higher Czech and Slovak rates compared with Germany are statistically attributable to platforms. Lists derived from Common Crawl miss about a third of the Slovak opt-outs because they exclude sites that block its crawler. The results describe delegated consent, a signal that a platform sets on behalf of its users, which measurements of consent on the web must consider.
\end{abstract}


\section{Introduction}\label{sec:intro}

Research that uses web data rarely asks who produced the signals it records. Many rules of the digital economy rely on consent signals that individuals are expected to set themselves, such as cookie choices, privacy preference signals and, most recently, opt-outs from the training of artificial intelligence (AI) models. Research on consent management shows that such signals are often produced by intermediaries whose interfaces and defaults determine what most users declare \citep{utz2019,nouwens2020,hils2020}. The same pattern may apply to the content on which AI models are trained. The European Union permits text and data mining of lawfully accessible content unless the rightholder has reserved this use ``by machine-readable means'' \citep[Art.~4(3)]{dsm2019}. In legal terms this opt-out is a reservation of rights. The AI Act obliges providers of general-purpose AI models to identify and respect it \citep[Art.~53(1)(c)]{aiact2024}. In practice the opt-out signal is a line in a robots.txt file or a metadata tag. Content management systems, e-commerce platforms, hosting providers and content delivery networks often write such lines for thousands of sites at once \citep[p.~5]{keller2026}.

The rule assumes that an opt-out expresses the decision of the rightholder. This paper uses the informed choice of the rightholder as a normative benchmark. It introduces the concept of delegated consent for a consent signal that a platform or another intermediary sets on behalf of its users. Based on research on defaults and consent intermediaries, platformisation, measurement frames and digital inequality (Section~\ref{sec:framework}), it formulates four expectations. Vendor defaults should produce many of the opt-outs of a national web. Differences between countries should follow their platform markets. Lists derived from crawls should undercount opt-outs, while popular sites should opt out more often than ordinary ones. These expectations are stated as four exploratory hypotheses and examined with descriptive analyses and logistic models.

Existing measurements cannot test these hypotheses because they use lists of top sites ranked by popularity, the domains included in AI training corpora or curated sets of news publishers \citep{longpre2024,liu2025,bouchaud2025,hoffmann2026,fletcher2024}. Such samples overrepresent large, professionally managed sites, which opt out far more often than the rest \citep{longpre2024}. \citet{dinzinger2024long} trace robots.txt files to content management systems, but no study separates the settings of vendors from the files written by site owners or compares countries that share the same legal rule.

This paper measures five national web zones that share the Directive but differ in language, in their national implementation of Article~4(3) and in their platform markets: Poland, the Czech Republic, Slovakia, Hungary and Germany. The complete lists of domains that Common Crawl recorded for these zones were crawled in October 2026, which gave 3,445,751 reachable sites. Because such lists cover only part of each zone, a random sample of 10,000 domains drawn from the public list of the Slovak registry serves as a benchmark. Every signal that can express an opt-out from AI training was collected and classified by its form, its scope and its provenance, which records whether a vendor or the site owner set it. The Tranco list \citep{lepochat2019tranco} was used to compare ranked and unranked sites. Four research questions guide the analysis:

\begin{enumerate}
\item[\textbf{RQ1}] How common are opt-outs from AI training on national webs?
\item[\textbf{RQ2}] In which legal and technical form are they expressed?
\item[\textbf{RQ3}] Who sets the opt-out signal: the site owner or a platform or hosting provider?
\item[\textbf{RQ4}] How do the domain list and the popularity of sites affect the measured rate?
\end{enumerate}

The paper makes three contributions to the study of consent on the web. First, it makes a methodological contribution to research with web data. It shows how consent signals can be attributed to the party that set them. It also measures how lists derived from Common Crawl exclude the sites that block its crawler, a coverage error in the sense of the total error framework for digital traces \citep{sen2021}, which extends work on the quality of digital behavioural data \citep{daikeler2025,weiss2025,dahlke2023} to the frames from which web data are drawn. Second, it develops the concept of delegated consent and extends research on defaults and consent intermediaries from privacy to the use of web content for AI. Where a dominant platform sets an opt-out by default, that single decision produces most opt-outs of a national web, about three quarters on the Czech and Slovak lists. Third, it provides a classification of the signals by form and provenance with a public dataset \citep{dataset2026}.

\section{Conceptual framework}\label{sec:framework}
The opt-out of Article~4(3) assumes that each rightholder decides whether to opt out. Four strands of research suggest that practice departs from this model in predictable ways. The result is delegated consent, a signal that expresses the setting of an intermediary unless the user changes it.

The first strand concerns defaults and consent intermediaries. Behavioural research shows that defaults influence choices strongly \citep{johnson2003} because the existing arrangement ``tends to stick'' \citep[p.~176]{thaler2003}. Studies of platform settings ask whether users configure them actively or show ``settings blindness'' \citep{haji2020}. Research on privacy consent offers the closest parallel. Field studies of consent notices find that preselected options, nudging and the position of a notice change how users respond \citep{utz2019}. Consent management platforms spread quickly across the web \citep{hils2020}, most of their notices fail minimal legal requirements \citep{nouwens2020} and some register consent before the user has chosen \citep{matte2020}. In social research, consent to share digital traces is itself selective and determines who enters the data \citep{gaughan2026}. Earlier machine-readable preference signals such as P3P, Do Not Track and Global Privacy Control show that the effect of a signal depends on whether its recipients accept it as the choice of the user \citep{hils2021}. Where a platform writes robots.txt for its customers, an opt-out set by default should therefore persist on most of its sites, while an opt-out that owners must switch on should remain rare.

The second strand concerns platform and infrastructural governance. Platforms set rules for their users through technical settings \citep{gorwa2019}. The platformisation of the web extends platform infrastructure into external websites \citep{helmond2015}, while producers that depend on a platform adapt their practices to its rules \citep{nieborg2018}. Infrastructure studies describe how a few providers come to determine practices across large parts of the web \citep{plantin2018}. Shops on a hosted e-commerce platform are such dependent producers, because the platform writes their robots.txt. Platform markets are often regional, as e-commerce platforms serve the language areas in which they operate. Differences between countries should then follow the geography of these markets.

The third strand concerns the frames used to measure the web. Research on digital behavioural data shows that the measurement process, including the choice of platform and data source, explains much of the variation in results \citep{schneck2024,weiss2025}. Total error frameworks for digital traces separate errors of representation, such as coverage, from errors of measurement \citep{sen2021,daikeler2025}. Content collected by scraping is systematically incomplete when pages cannot be reached \citep{dahlke2023}. The conditions of access also change what researchers receive \citep{holzl2026}, while top lists used in web research are unstable and biased towards particular kinds of sites \citep{scheitle2018}. Common Crawl, the main source of web data for training AI models, collects only pages that its crawler is allowed to fetch \citep{baack2024}. Lists of domains derived from it therefore exclude the sites that block its crawler. Because blocking a crawler is itself an opt-out, such lists should undercount the behaviour that they are used to measure. Popularity lists in turn overrepresent large, professionally managed sites, which opt out more often \citep{longpre2024}.

The fourth strand concerns inequality in the use of digital technologies. Research on the second level of the digital divide shows that differences in skills and usage persist once access is common \citep{hargittai2002,vandeursen2014,scheerder2017}. Digital inclusion is gradual and actors with fewer resources are more often excluded from its benefits \citep{ragnedda2022}. Applied to AI training, the capacity to opt out should depend on resources, so professionally managed sites should opt out more often than ordinary ones. Protective behaviour online, however, depends less on literacy than on attitudes such as privacy cynicism \citep{respi2026}. Popularity is moreover only an imperfect indicator of resources. Four hypotheses follow:

\begin{enumerate}
\item[\textbf{H1}] Sites of platforms that set an opt-out by default opt out far more often than other sites, while an opt-out that owners must switch on remains rare.
\item[\textbf{H2}] Differences between countries shrink once the platform of each site is taken into account.
\item[\textbf{H3}] Sites missing from the lists derived from Common Crawl opt out more often than listed sites because of opt-outs that name the crawler of Common Crawl.
\item[\textbf{H4}] Sites on a popularity list opt out more often than unranked sites, also when platform and country are taken into account.
\end{enumerate}

\noindent H1 and H2 test delegated consent and concern RQ3, while H3 and H4 concern RQ4. RQ1 and RQ2 are descriptive and provide the basis for testing them. The hypotheses were formulated after a first descriptive analysis of the data and are therefore exploratory. The analyses examine whether the observed patterns persist under adjustment and alternative specifications. H2 is examined as a decomposition of country differences and H3 as an estimate of coverage bias.

\section{Technical context}\label{sec:background}
The legal rule, its national implementations and the related case law are summarised in Appendix~\ref{app:legal}. Three kinds of signals can express a reservation. The Robots Exclusion Protocol \citep{rfc9309} instructs crawlers by name, so a robots.txt file can disallow known AI crawlers such as GPTBot or CCBot, the crawler of Common Crawl. The TDM Reservation Protocol (TDMRep) expresses a reservation in a file, an HTTP header or a meta tag \citep{tdmrep2024}. Content Signals, introduced by Cloudflare in 2025, add a line such as \texttt{Content-Signal: ai-train=no} to robots.txt \citep{cloudflare2025}.

\begin{figure}[htbp]
\centering
\includegraphics[width=\textwidth]{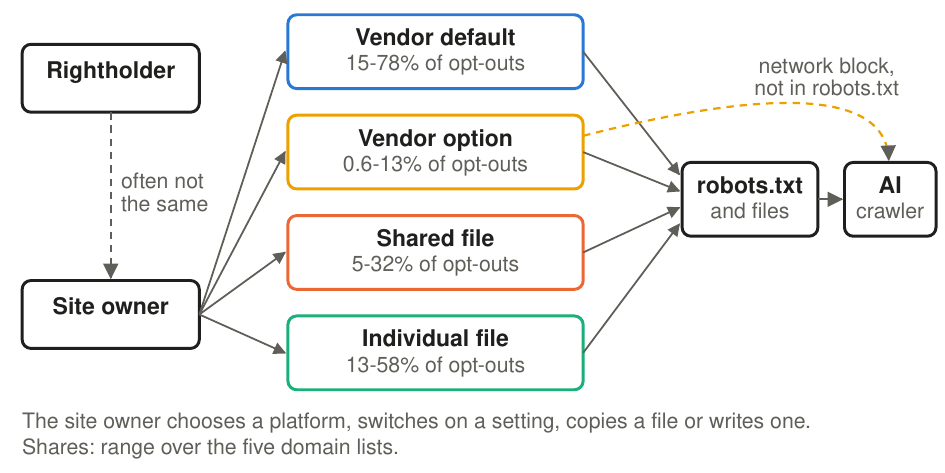}
\caption{Who sets the opt-out signal that an AI crawler reads}
\label{fig:chain}
\end{figure}

Most websites do not write their own robots.txt. Content management systems, e-commerce platforms, hosting providers and content delivery networks generate or modify it. Site owners are often not the rightholders of the content they publish \citep[p.~10]{keller2026}. Vendors set opt-outs in robots.txt in different ways. Cloudflare offers a managed robots.txt that customers switch on \citep{cloudflare_managed_docs} and since July 2025 has asked new domains whether to block AI crawlers in its network \citep{cloudflare2025default}. Squarespace offers a setting that is off unless the owner selects it \citep{squarespace_help2026}. The Polish platform IdoSell documents a switch that a merchant must set to yes for robots.txt to admit AI robots \citep{idosell_help2026}. Figure~\ref{fig:chain} summarises who writes the file that an AI crawler reads, with the shares of opt-outs measured in Section~\ref{sec:res:choice}.

\section{Related work}\label{sec:related}

\citet{longpre2024} audit the domains of three AI training corpora and find restrictive robots.txt rules on 26.5\% of the top 2,000 domains by token count against 3.4\% of a random sample of the corpus domains. \citet{liu2025} study popular sites on the Tranco list, survey artists and show that platform features such as the blocking option of Squarespace are relevant for creators. \citet{bouchaud2025} find GPTBot fully blocked on 10.6\% of the top one million sites in the Chrome browser data. \citet{hoffmann2026} fetch robots.txt, \texttt{ai.txt} and \texttt{llms.txt} from the 4.08 million hosts of the Tranco list. \citet{dinzinger2024survey,dinzinger2024long} use Common Crawl data, find a TDMRep file on only 45 hosts and estimate that at least 10\% of robots.txt files are generated by content management systems or tools for search engine optimisation. \citet{fletcher2024} reports that 48\% of the top news sites of ten countries blocked the crawlers of OpenAI by the end of 2023. \citet{steinacker2026} find that 60.0\% of reputable news sites with a robots.txt disallow at least one AI crawler against 9.1\% of misinformation sites, a gap that popularity alone does not explain. Studies of compliance by crawlers \citep{kim2025,liu2025,lopez2026} measure what crawlers do, whereas this study measures what sites declare. Policy analysis describes how infrastructure providers influence this practice \citep{keller2026}.

None of these studies uses complete national domain lists with a registry benchmark, separates vendor settings from files written by owners or compares countries under the same directive.

\section{Data and methods}\label{sec:methods}

\subsection{Domain lists and registry sample}
For each zone the domains recorded in the Common Crawl crawls of July to September 2026 were imported and every domain on the list was crawled. The Polish list was extended with nine earlier crawls and with link discovery. The lists cover 13.4\% of the registered domains in \texttt{.de}, 18.3\% in \texttt{.hu}, 22.7\% in \texttt{.cz}, 23.7\% in \texttt{.sk} and 31.0\% in \texttt{.pl} \citep{reg_denic,reg_domainhu,reg_cznic,reg_sknic,reg_dnspl}. Because the lists were crawled completely, their estimates carry no sampling error but may be biased by coverage. To measure this bias, 10,000 domains were drawn at random from the list of all 493,505 registered \texttt{.sk} domains, which the Slovak registry publishes \citep{reg_sknic}. Of these, 2,254 were on the Common Crawl list and kept their results, while the other 7,746 were crawled on 25~October 2026 with the same crawler. The other registries do not publish their lists. The unit of analysis is the registrable domain, the domain that a registrant can register under the zone. The main denominator is the set of reachable domains, which answered at least one request.

\subsection{Collection and classification}
For each domain the robots.txt file, the headers and meta tags of the home page, the TDMRep file and the files \texttt{llms.txt} and \texttt{ai.txt} were collected. The crawler honoured robots.txt, applied the Crawl-delay of each site up to ten seconds and used no login. A site counts as opting out if it shows a signal in one of three tiers of legal form. Tier~1 is a protocol built for text and data mining (TDMRep). Tier~2 is a machine-readable directive or signal: a robots.txt group that names a generative AI crawler from a curated list and disallows the whole site, the line \texttt{Content-Signal: ai-train=no} or a \texttt{noai} tag. Tier~3 is a statement in natural language in robots.txt comments, \texttt{llms.txt} or \texttt{ai.txt} without a machine-readable equivalent. Groups for all crawlers that disallow the whole site name no AI use and are not counted, but their prevalence is reported. The scope of each opt-out is training only or training together with search and assistants, following the values of Content Signals and the classes of the blocked crawlers.

The provenance records who wrote the robots.txt of the site. A vendor default is a file that matches the template of a platform or hosting provider that writes the opt-out unless the owner changes it (Shoptet, IdoSell and the German hosting provider PixelX). A vendor option is a template whose opt-out the owner switches on (the managed robots.txt of Cloudflare and the setting of Squarespace). The setting of each template was established from vendor documentation, archived files and the files themselves (Section~\ref{sec:res:choice}). A shared file matches no template and is identical byte for byte on at least two sites with an opt-out, which indicates a framework, plugin or copied list. All other opt-outs are classified as individual files written for one site.

\subsection{Validation}
The classification was compared with blind labels on a stratified holdout sample of 300 sites (60 per zone), weighted back to the population by stratum. A fresh sample of 49 statements in natural language was drawn after the rules for that tier were revised. Recall was checked by labelling sites without an opt-out that contain words an opt-out would need. Labels were produced blind by a language model following a written coding guide, without access to the classifier output. Terms of service were examined in a separate random sample of 8,000 sites per zone with a link from the home page to a terms, regulations or legal notice page on the same site. Up to two such pages per site were fetched with the same rules as the main crawl and scanned for opt-outs from text and data mining or AI training. All flagged pages and 20 unflagged pages per zone that mention these topics were labelled blind.

\subsection{Statistical analysis}
Because the lists were crawled completely and a few provenance sources carry many sites, results for the lists are reported as descriptive values of the listed population. Ratios of rates in Figure~\ref{fig:toplist} carry Katz log intervals as a descriptive measure of their precision. Inference is confined to the registry sample, which carries Wilson score intervals for proportions. The hypotheses are examined with logistic regressions of the presence of an opt-out on reachable sites. Model~1 contains the country of the site as its only predictor. Model~2 adds the platform identified from the robots.txt template (Shoptet, IdoSell, Squarespace or none). Model~3 also adds the use of Cloudflare and the Tranco rank in three bands, with unranked sites as reference. Model~4 uses the Slovak registry sample and relates opting out to absence from the list, the platform and Cloudflare. Model~5 repeats it with opt-outs that do not name CCBot as the outcome. Intervals for Models~4 and~5 come from a cluster bootstrap over 535 providers of name servers (999 resamples), which group sites by hosting independently of the outcome. Shoptet shops are identified by the paths of their cart and filters, IdoSell shops by a comment of their recommendation system and Squarespace sites by the header of its file. The identification thus uses no opt-out line and every platform includes sites without an opt-out. The 2,478 sites of PixelX are excluded from the models, because this provider is recognised only by the file that carries the opt-out. Templates without a setting for AI crawlers count as no platform. Because odds ratios of nested logistic models are not comparable \citep{mood2010}, differences between countries are compared across models as average marginal effects in percentage points against Germany, with linear probability models as a check. Since the platform is identified from the file that also carries the opt-out, the models measure associations and do not identify causal effects. All variables are part of the published dataset.

\section{Results}\label{sec:results}

\subsection{Prevalence of opt-outs (RQ1)}\label{sec:res:rq1}
\begin{figure}[htbp]
\centering
\includegraphics[width=\textwidth]{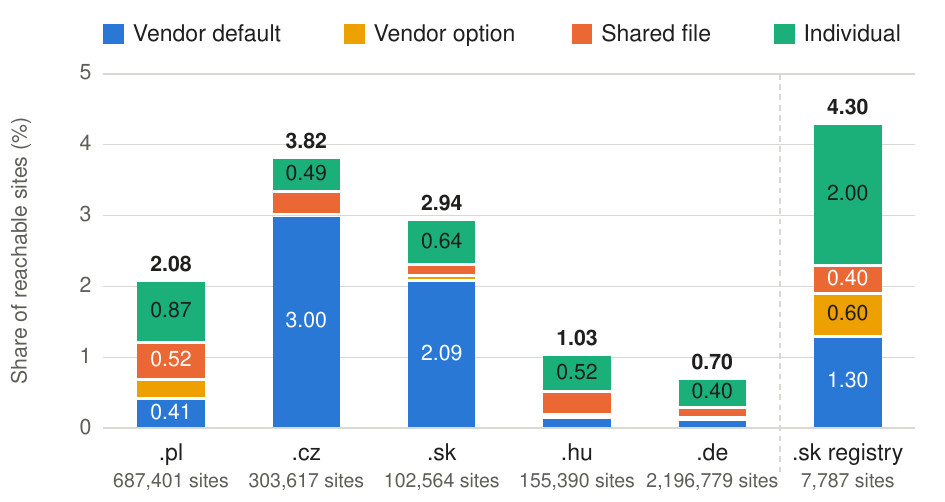}
\caption{Opt-out rate by zone and provenance in per cent of reachable sites}
\label{fig:provenance}
\end{figure}

Figure~\ref{fig:provenance} shows the share of reachable sites with an opt-out in each zone, divided by provenance, with the number of reachable sites under each bar. Of the 3,445,751 reachable sites on the lists, 45,837 publish an opt-out signal, from 0.70\% in \texttt{.de} to 3.82\% in \texttt{.cz}. The last bar gives the Slovak registry sample, in which 4.30\% of the 7,787 reachable sites opt out, with a 95\% confidence interval (CI) of 3.87\% to 4.78\%. Its four provenance groups correspond to 101, 47, 31 and 156 sites. Because of rounding the segments may not add up to the total.

\subsection{Legal and technical form (RQ2)}\label{sec:res:form}
Crawler directives in robots.txt and Content Signals are the primary mechanism of more than 93\% of opt-outs in every zone. Crawler directives dominate in \texttt{.pl}, \texttt{.hu} and \texttt{.de}, while Content Signals account for 81\% of opt-outs in \texttt{.cz} and 75\% in \texttt{.sk}. TDMRep is the primary mechanism for at most 1.73\% of opt-outs and TDMRep files with an opt-out were found on 1 to 173 domains per zone. Statements in natural language without a machine-readable signal account for 0.41\% to 2.47\% of opt-outs, except in \texttt{.sk}, where an \texttt{ai.txt} file deployed by a vendor on many sites raises the share to 4.51\%. Terms of service, which tier~3 does not cover, were examined separately. Opt-outs in terms pages concern an estimated 0.005\% (\texttt{.sk}) to 0.092\% (\texttt{.pl}) of reachable sites. The labels confirmed 54 of the 59 flagged pages. Even in \texttt{.de}, where 67.8\% of sites link to a terms page, opt-outs found only in terms pages would add about 5\% to the opt-outs, with at most 12\% at the upper bound of the interval. Most crawler directives address individual crawlers by their names. Among sites that opt out with a crawler directive, 18\% (\texttt{.hu}) to 53\% (\texttt{.pl}) also block CCBot. Between 20\% (\texttt{.pl}) and 42\% (\texttt{.sk}) of them name the crawlers of a single operator. Opt-outs limited to training make up 93\% in \texttt{.cz} and 86\% in \texttt{.sk}, where Content Signals set \texttt{ai-train=no} only. In the other zones they make up 45\% to 75\%, because many files also block AI search and assistant crawlers.

\subsection{Provenance of opt-outs (RQ3)}\label{sec:res:choice}
On the lists vendor defaults account for 78\% of opt-outs in \texttt{.cz} and 71\% in \texttt{.sk} but only for 15\% to 20\% in the other zones. Vendor options account for 13\% of opt-outs in \texttt{.pl}, mainly through the managed robots.txt of Cloudflare. In the other zones they account for at most 3\%. Individual files are the most frequent provenance in \texttt{.pl}, \texttt{.hu} and \texttt{.de} with 42\%, 51\% and 58\% of opt-outs, as well as in the Slovak registry sample with 47\%. Figure~\ref{fig:provenance} shows the same groups in per cent of reachable sites.

\begin{figure}[htbp]
\centering
\includegraphics[width=\textwidth]{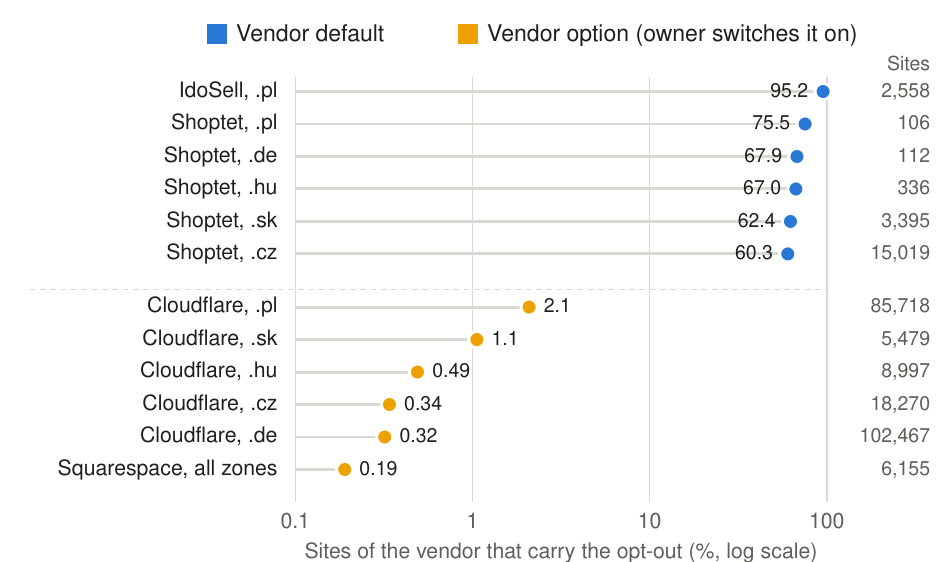}
\caption{Share of each vendor's sites that carry the opt-out on a logarithmic scale}
\label{fig:uptake}
\end{figure}

Figure~\ref{fig:uptake} shows the share of each vendor's sites that carry the opt-out, which indicates how the vendor settings work. Of the 2,558 Polish IdoSell shops, identified by the paths of their robots.txt, 95.2\% carry the block of AI crawlers, which agrees with a setting that is active unless the merchant changes it. Of the shops whose robots.txt has the structure of the Shoptet template, 60.3\% in \texttt{.cz}, 62.4\% in \texttt{.sk} and 67.0\% in \texttt{.hu} carry \texttt{Content-Signal: search=yes, ai-train=no}. About 1\% carry a line that permits all uses and the others have no Content-Signal line. The files of the three groups are otherwise identical, so the line is the value of a platform setting with at least three states. The few Shoptet shops in \texttt{.pl} and \texttt{.de} show similar shares (Figure~\ref{fig:uptake}). Archived copies in Common Crawl show that the line appeared on 23 of 25 sampled shops between their captures of 7 to 18~August and 4 to 17~September 2026, which agrees with a single release of the platform. The help pages of Shoptet describe three settings for search, AI answers and AI training, each with the values allow, disallow or not set. Their default allows search, leaves AI answers unset and disallows training \citep{shoptet_help2026}, which produces exactly the observed line. A trial shop opened on 5~November 2026 showed these preselected values in its administration. The platform writes a signal only when its value is set \citep{zahalka2026}, so the opt-out line is a default that owners can change. Its position in the part of the file that the platform writes agrees with this \citep{custela2026}. Of the sites served through Cloudflare, only 0.32\% to 2.09\% use its managed robots.txt. Of the live Squarespace sites 0.19\% use its setting for AI crawlers. Both agree with settings that customers switch on. Settings that are active by default thus reach 60\% to 95\% of a platform's sites, while settings that owners switch on reach at most 2.1\%. The hosting provider PixelX serves its file to every hosted site without a robots.txt of its own. Its sites are recognised only by this file, so the share of its sites that carry the opt-out cannot be observed.

\begin{figure}[htbp]
\centering
\includegraphics[width=\textwidth]{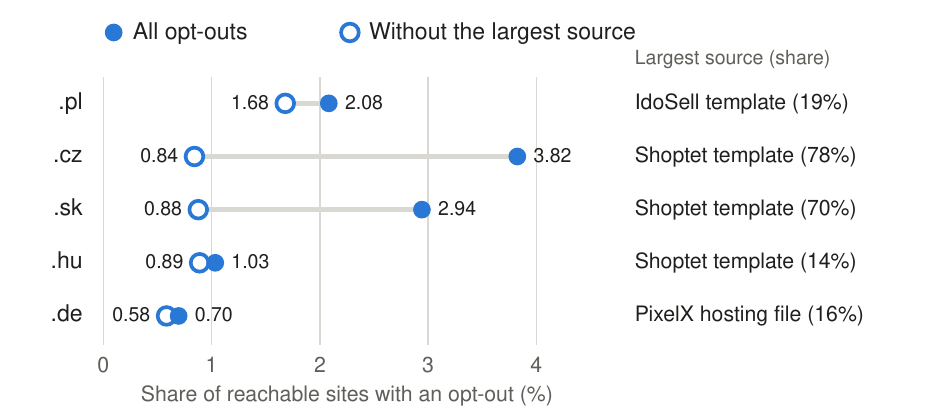}
\caption{Opt-out rates with and without the largest source of opt-outs in each zone}
\label{fig:counterfactual}
\end{figure}

Figure~\ref{fig:counterfactual} shows how concentrated the sources of opt-outs are on the lists. For each zone it gives the share of reachable sites with an opt-out and the share that would remain without the largest source of opt-outs. A source is a template, a shared file or, for individual files, the site itself. The comparison assumes that the sites of the largest source would publish no opt-out without it. In \texttt{.cz} and \texttt{.sk} one template produces 78\% and 70\% of all opt-outs. Without it, the share of sites with an opt-out would fall from 3.82\% to 0.84\% and from 2.94\% to 0.88\%, within the range of the other zones without their largest source (0.58\% to 1.68\%). The three largest sources together account for 19\% (\texttt{.de}) to 81\% (\texttt{.cz}) of the opt-outs. Individual files remain upper bounds for files written by owners, because 43\% of them were labelled as generated. When files are compared after removing sitemap lines, the domain of the site, case and blank lines, the share of reachable sites with an individual file falls to 0.23\% to 0.44\%. Files written by owners themselves are therefore fewer than Figure~\ref{fig:provenance} suggests.

\subsection{Effect of the sampling frame (RQ4)}\label{sec:res:frame}
\begin{figure}[htbp]
\centering
\includegraphics[width=\textwidth]{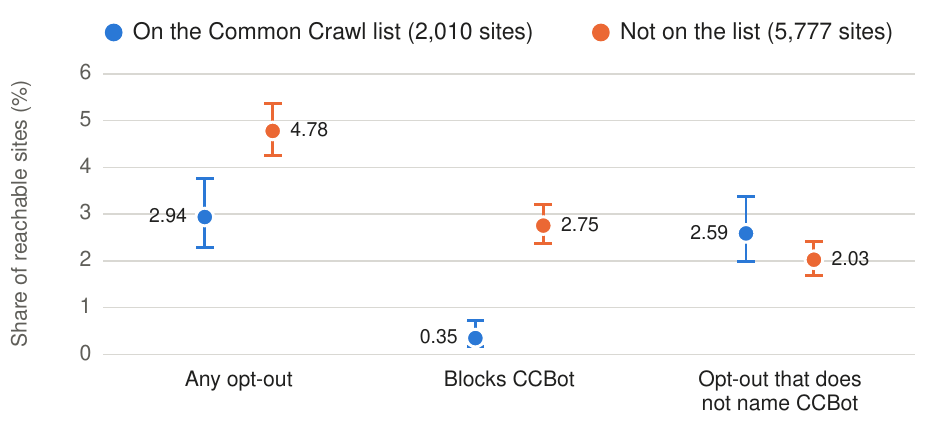}
\caption{Opt-out rates in the Slovak registry sample on and off the Common Crawl list}
\label{fig:frame}
\end{figure}

The registry sample shows that the lists derived from Common Crawl undercount opt-outs (Figure~\ref{fig:frame}). Within the registry sample, the 2,010 reachable sites on the Common Crawl list opt out at 2.94\%, the same rate as the full \texttt{.sk} list after rounding. The 5,777 reachable sites that are not on the list opt out at 4.78\% (95\% CI 4.26-5.36\%). The difference comes from sites that block CCBot. They make up 2.75\% of the sites off the list against 0.35\% of the sites on it (Fisher's exact test, $p<10^{-12}$). Common Crawl honours robots.txt, so a site that blocks its crawler rarely enters its index and therefore rarely enters a list derived from it. Opt-outs that do not name CCBot occur at similar rates on and off the list (2.59\% and 2.03\%, $p=0.15$). The lists also differ in their provenance mix. In the registry sample vendor defaults account for 30\% of opt-outs and individual files for 47\%, against 71\% and 22\% on the \texttt{.sk} list. Vendor options are more frequent in the registry sample because 46 of its 47 sites with the managed robots.txt of Cloudflare are off the list, as this file blocks CCBot. The same pattern appears within the Common Crawl lists. In \texttt{.cz}, \texttt{.sk} and \texttt{.hu} sites recorded in all three crawls opt out more often than sites recorded once, mainly because of platform shops (4.95\% against 1.04\% in \texttt{.cz}). On the part of the Polish list drawn from the same three crawls as the other lists, 1.32\% of sites opt out. Of the domains found through links 7.78\% opt out, as do 3.40\% of those from earlier crawls. The coverage bias of the list is thus 1.36 percentage points, about a third of the opt-outs in the zone. The estimates for the five lists are therefore likely lower bounds for the share of sites with an opt-out in each zone and overstate the share of vendor defaults. The Polish rate is not directly comparable with the others.

\subsection{Popularity bias (RQ4)}
\begin{figure}[htbp]
\centering
\includegraphics[width=\textwidth]{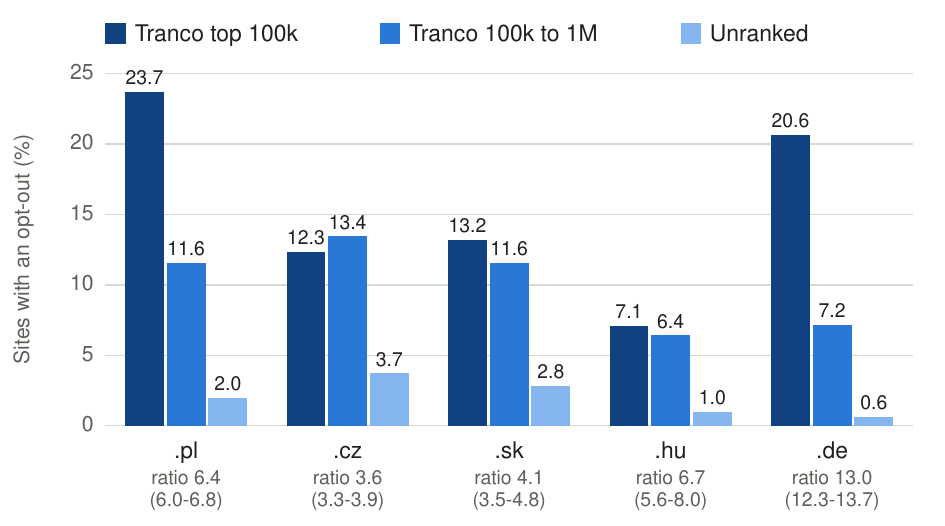}
\caption{Opt-out rate by Tranco rank and zone}
\label{fig:toplist}
\end{figure}

Figure~\ref{fig:toplist} compares reachable sites on the Tranco list with unranked sites and gives the ratio of their rates under each zone. Ranked sites are 0.78\% to 1.20\% of each zone but opt out 3.6 to 13.0 times more often than unranked sites. The ratio is lowest in \texttt{.cz} and \texttt{.sk} because the Shoptet shops are rarely ranked. Without template files it lies between 6.5 and 15.9. In every zone except \texttt{.cz} the rate is highest in the top 100,000 and 31.3\% of the 131 German sites in the top 10,000 opt out. Popular sites and the national web are therefore different populations.

\subsection{Tests of the hypotheses}\label{sec:res:models}
Table~\ref{tab:models} reports the models for the five lists as descriptive values. Sites of platforms that opt out by default have far higher odds of opting out than sites without a known platform. The odds of Squarespace sites, whose setting the owner must switch on, are only 2.8 times higher in Model~2. Of the sites matched by the templates, 94.8\% of IdoSell shops and 60.9\% of Shoptet shops opt out against 1.8\% of the sites with a Squarespace file, which agrees with H1. These shares differ from Figure~\ref{fig:uptake}, which identifies IdoSell shops by the paths of their files and counts only live Squarespace sites, while the Squarespace template also matches files of unpublished sites. Adding the platforms reduces the average marginal effect of the Czech zone against Germany from 3.24 to 0.33 percentage points and that of the Slovak zone from 2.36 to 0.38 points. In this decomposition platforms account for 90\% and 84\% of these differences (92\% and 85\% in linear probability models), as H2 expects. The Polish effect falls only from 1.50 to 1.24 points, since individual files and the managed robots.txt of Cloudflare account for more of its opt-outs. Adjustment for Cloudflare and the Tranco rank in Model~3 changes the country effects little. Ranked sites have 8.3 to 58.5 times the odds of unranked sites in Model~3, against 8.4 to 35.7 times in a model with the Tranco rank alone. The gap thus persists after adjustment and widens for the top bands, which agrees with H4.

\begin{table}[htbp]
\centering
\setlength{\tabcolsep}{6pt}
\caption{Logistic models of opting out on the five lists}
\label{tab:models}
\footnotesize
\begin{tabular}{@{}l r r r@{}}
\toprule
 & Model 1 & Model 2 & Model 3 \\
\midrule
\multicolumn{4}{@{}l}{\emph{Odds ratios: country (reference DE)}} \\
CZ & 6.77 & 1.47 & 1.39 \\
HU & 1.78 & 1.55 & 1.46 \\
PL & 3.62 & 2.95 & 2.47 \\
SK & 5.16 & 1.56 & 1.48 \\
\multicolumn{4}{@{}l}{\emph{Odds ratios: platform (reference none)}} \\
Shoptet &  & 181 & 223 \\
IdoSell &  & 1,189 & 1,554 \\
Squarespace &  & 2.77 & 3.34 \\
\multicolumn{4}{@{}l}{\emph{Odds ratios: other predictors}} \\
Cloudflare &  &  & 3.55 \\
Tranco 100k to 1M &  &  & 8.28 \\
Tranco 10k to 100k &  &  & 23.6 \\
Tranco top 10k &  &  & 58.5 \\
\multicolumn{4}{@{}l}{\emph{Average marginal effects against DE (percentage points)}} \\
CZ & 3.24 & 0.33 & 0.28 \\
HU & 0.45 & 0.38 & 0.33 \\
PL & 1.50 & 1.24 & 0.97 \\
SK & 2.36 & 0.38 & 0.34 \\
\midrule
Reachable sites & 3,443,273 & 3,443,273 & 3,443,273 \\
McFadden $R^2$ & 0.054 & 0.242 & 0.272 \\
\bottomrule
\end{tabular}
\end{table}

Table~\ref{tab:registrymodels} reports the models of the registry sample with bootstrap intervals. Sites missing from the list have 2.40 times the odds of opting out once platform and Cloudflare are taken into account (Model~4). When opt-outs that name CCBot are excluded from the outcome, the odds ratio falls to 1.21 with an interval that includes one (Model~5). The difference between the frames is thus due to sites that block the crawler of Common Crawl, as H3 expects. This result is close to definitional, since a crawler that is blocked cannot record the site, but it measures a coverage bias that changes the observed rate in the Slovak zone by almost half.

\begin{table}[htbp]
\centering
\setlength{\tabcolsep}{3pt}
\caption{Logistic models of opting out in the Slovak registry sample}
\label{tab:registrymodels}
\footnotesize
\begin{tabular}{@{}l l l@{}}
\toprule
 & Model 4 (any opt-out) & Model 5 (not naming CCBot) \\
\midrule
Missing from the list & 2.40 (1.27 to 3.23) & 1.21 (0.73 to 1.65) \\
Shoptet & 84.9 (51.1 to 122) & 253 (115 to 463) \\
Cloudflare & 8.73 (2.32 to 12.5) & 8.10 (unstable) \\
\midrule
Sites with an opt-out & 335 & 169 \\
McFadden $R^2$ & 0.242 & 0.415 \\
\bottomrule
\end{tabular}
\end{table}

\subsection{Robustness and validation}
The order of the zones stays the same when crawler directives that name none of the seven major AI operators are excluded and when blocks of other AI crawlers are counted. Counting blocks of all crawlers raises the shares to between 1.19\% and 4.03\%. Opt-outs whose only signal is a block of CCBot, which also collects data for research and archiving, make up 0.5\% to 7.2\% of the opt-outs. Without them the rates change by at most 0.15 points (\texttt{.pl}: 1.93\% against 2.08\%). On the part of the Polish list drawn from the same three crawls as the other lists, the Polish odds ratio in Model~3 falls from 2.47 to 1.30. Its marginal effect falls from 0.97 to 0.21 points, so most of the remaining Polish difference comes from the domains added to its list. A model of the sites without a known platform gives country effects close to those of Model~3. In the registry sample, 40 reachable domains (0.5\%) are delegated to parking services. Without them the rate off the list is 4.81\% against 4.78\% with them. Parking pages hosted by ordinary registrars cannot be identified from the name servers. The holdout sample contained 121 sites that the classifier marked as opting out and 179 that it did not, because the strata oversampled sites with an opt-out. The labels agreed on the presence of an opt-out for 299 of the 300 sites, with one false positive and no missed opt-out. Precision is thus 120 of 121 (99.2\%, Clopper-Pearson 95\% CI 95.5\% to 100\%), with full agreement on tier. The rules for statements in natural language were revised after the holdout showed a false positive in this tier. On a fresh sample the blind labels confirmed 40 of 49 such statements (weighted 89\%). The recall check scanned 20,000 sites without an opt-out per zone and labelled 100 sites in total from the 7.49\% that contained relevant words. It found one missed opt-out, a statement in an \texttt{ai.txt} file. With the upper 95\% Clopper-Pearson bound of 5.45\% for one miss in 100, missed opt-outs concern at most 0.41\% of reachable sites against 1.33\% with an opt-out, which bounds recall at 76\% or more. Under the revised rules, which detect this statement, the bound of 3.62\% for no miss gives 83\% or more. The check does not cover AI crawlers missing from the curated list.

\section{Discussion}\label{sec:discussion}

\subsection{Theoretical implications}
The results agree with the four exploratory hypotheses, with qualifications that follow from the design. An opt-out that a vendor sets by default reaches 60\% to 95\% of its sites, while one that owners must switch on reaches at most 2.1\% (Figure~\ref{fig:uptake}). This difference of more than an order of magnitude agrees with the stickiness of defaults (H1). The Czech and Slovak zones show nearly the same pattern under different national texts, because the same platform serves both markets. In a decomposition, 84\% to 90\% of their difference from Germany is attributable to platforms (H2). Elsewhere individual files are the most frequent provenance, so the influence of vendors depends on the structure of each platform market. Five countries do not allow the effect of the national texts to be separated from other national differences. The results therefore show that platform markets account for most of the observed differences without testing the law itself. The coverage bias of the frame (H3) is due to sites that block CCBot. Ranked sites opt out more often also after adjustment for platform and country (H4).

Taken together, the results describe delegated consent, since the opt-out was designed as an individual decision of each rightholder but in practice operates through infrastructure. This repeats a pattern known from privacy, where consent management platforms and their defaults determine most of the signals that users send \citep{nouwens2020,hils2020}. One difference concerns the direction of the default. Consent interfaces often favour the collection of data \citep{utz2019,mathur2019}, while the vendors studied here set defaults that favour the rightholder. A decision of one vendor applies to thousands of sites at once and works in both directions. The sites of platforms that write no opt-out signal remain open to AI crawlers without any decision of their owners. Shops on hosted platforms, whose robots.txt is part of the platform, are dependent producers in the sense of research on platformisation \citep{helmond2015,nieborg2018}. One setting designed and preselected by a single platform produces about three quarters of the opt-outs on two national lists. The mechanism described for the settings of platform users \citep{haji2020} thus also applies to a rule of copyright law. In this sense the providers of content management systems, e-commerce platforms and hosting services take part in the governance of the data used to train AI models.

Delegated consent may also have economic and cultural consequences, which the data do not measure. A change of one default can make thousands of small online shops visible or invisible to AI assistants. The defaults of a few platforms also change how much Czech, Slovak and Hungarian content remains available for training models in smaller languages.

The popularity gap agrees with research on digital inequality, in which actors with more resources use digital options more fully \citep{hargittai2002,scheerder2017,ragnedda2022}. The data measure neither resources nor preferences, however. Popularity may also reflect the value of content to AI developers or its exposure to public debate. Owners of small sites may decide not to opt out because they want to remain visible in AI search. Protective behaviour online follows attitudes such as privacy cynicism more than literacy \citep{respi2026}, so the absence of an action does not by itself indicate a lack of capacity. The gap shows that the opt-out is exercised unevenly, while its causes require data on site owners. For most small sites on hosted platforms the decision lies in practice with the platform, which can protect owners who would not act themselves but also decides for them.

\subsection{Methodological implications}
Research that uses web data should treat consent signals as attributable to the party that set them. Matching files against the templates of platforms and grouping identical files separate the settings of vendors from the decisions of owners and can be applied to other signals, such as \texttt{llms.txt} files or cookie banners. Delegated consent is in this sense a measurement error, since a signal attributed to a site may express the setting of a vendor. Crawl archives in turn are not neutral samples of the web. Like pages that scraping cannot reach \citep{dahlke2023}, sites that refuse a crawler are missing from every dataset built on it, a coverage error in the sense of the total error framework \citep{sen2021,weiss2025}. Studies that measure opting out, or the composition of training data, from lists derived from Common Crawl are therefore likely to underestimate refusal and to overstate the weight of large platforms. A random sample from a registry, even of one zone, shows the size of this bias.

\subsection{Implications for platform design and policy}
Whether a default counts as the decision of the rightholder is legally open \citep[p.~17]{olg2025}. With vendor defaults at 15\% to 78\% of the opt-outs, the answer decides the status of a large part of the opt-out. Platforms can choose among three designs of the opt-out setting. A default with a disclosure informs owners that the platform opts out for them and how to change it. A prompt during onboarding, such as the question that Cloudflare asks new domains, records an explicit decision. No default leaves the decision to owners, most of whom do not take it, as the low uptake of vendor options shows. The first design maximises the reach of the opt-out, the second its attribution to an informed choice and the third the autonomy of the owner. Authorities that monitor the opt-out should report vendor defaults separately from individual opt-outs and rely on random samples drawn from registries, because lists derived from Common Crawl exclude the sites that block its crawler.

\section{Limitations and future research}\label{sec:limits}
The data describe the state of October 2026. The five lists probably give lower bounds for their zones. The Polish list was built from more sources than the others and is not directly comparable with them. Only the Slovak registry permits a random sample, so the size of the frame bias in the other zones is unknown. Within the registry sample, domains on the list were crawled between five and sixteen days before the others. Parked domains could be identified only where they are delegated to parking services. Domains that redirect to the same site were not merged. Blocking of AI crawlers in the network of a content delivery network leaves no trace in the collected files, so the influence of intermediaries is if anything underestimated. Platforms are identified from robots.txt, so shops that replaced the file of their platform are counted as sites without a platform and the platform effects are likely overstated. Provenance relies on known templates and byte identity, which miss templates with local changes and may group short files written independently. The setting of PixelX is known only from its files. The defaults of Shoptet and IdoSell are documented by their help pages, a trial shop and the shares of their shops that carry the opt-out. The validation used labels of a language model, whose risks for annotation are documented \citep{lin2026}. The rules for natural language were revised after the holdout was labelled, which the fresh sample addresses. Tranco rank measures popularity, which is only an imperfect indicator of the resources of site owners. The classification makes no statement on legal validity, while compliance by crawlers is outside the scope of the study.

Future research can build on the dataset in five directions. Surveys or interviews of site owners on hosted platforms would show whether they know the setting that determines their opt-out and would answer the question of whose choice it is. Repeated waves of the same collection would show how opt-outs spread and how quickly vendors change their defaults. Changes of a default, such as the introduction of Content Signals by a platform, offer natural experiments on its effect. Random samples from the registries of the other zones would measure the frame bias beyond Slovakia. Experiments with sites under the control of researchers could test which forms of opt-out are respected by crawlers in practice.

\section{Conclusion}\label{sec:conclusion}
The opt-out from AI training is designed as a decision of each rightholder. On the national webs of Poland, the Czech Republic, Slovakia, Hungary and Germany, however, it is largely delegated consent set by platform infrastructure. Opt-outs are rare and are expressed almost entirely through robots.txt and Content Signals. Where a dominant platform sets an opt-out by default, as on the Czech and Slovak lists, its setting produces most of the opt-outs of a national web, so national differences follow regional platform markets. The Slovak registry sample shows that lists derived from Common Crawl exclude the sites that block its crawler, so they undercount opt-outs and overstate the share of vendor defaults. Popular sites opt out far more often than the rest, so the opt-out is used unevenly. The design of platform defaults, the monitoring of the opt-out and the interpretation of Article~4(3) therefore have to take into account who sets the signal and which frame is used to measure it.

\section*{Ethics statement}
The study collected only files that any visitor of a website can request: robots.txt, the home page, linked terms pages and three files at fixed locations. The crawler followed robots.txt, applied the Crawl-delay of each site up to ten seconds, used no login or form and identified itself with a fixed user agent. No human participants took part in the study. Domains of natural persons can be personal data, as can other content collected from the web \citep{rohatgi2025}, so the data are processed for research purposes only. From the public list of the Slovak registry only domain names and name servers were kept. Processing for scientific research follows the safeguards of Article~89 of the General Data Protection Regulation, in particular data minimisation and the pseudonymisation of the released data. As no human participants were involved, the study did not undergo review by an ethics committee. The released dataset contains parsed fields without texts of third parties or addresses of terms of service pages.

\section*{Declaration of conflicting interests}
The author declares no potential conflicts of interest with respect to the research, authorship or publication of this article.

\section*{Funding}
The author received no financial support for the research, authorship or publication of this article.

\section*{Data availability statement}
The parsed signals and the classification of each listed site are available as version~1.0 of October 2026 on Zenodo under a CC~BY~4.0 licence at \url{https://doi.org/10.5281/zenodo.23686009} \citep{dataset2026}. The repository also contains the results of the Slovak registry sample and of the terms of service sample, with domain names replaced by keyed hashes. Raw texts of third parties, headers that reveal server versions, addresses of terms of service pages and fields of the registry other than the domain are withheld for legal and data protection reasons. The popularity ranking used is Tranco list N26LW \citep{lepochat2019tranco}. The code used for the collection and the analysis is available from the author on request.

\appendix
\renewcommand{\thetable}{A\arabic{table}}
\setcounter{table}{0}
\section{Regulatory context, national implementation and case law}\label{app:legal}
\subsection*{Regulatory context}
Article~4 of the Directive on copyright in the Digital Single Market \citep{dsm2019} permits anyone, including commercial developers of AI models, to reproduce lawfully accessible works for text and data mining unless their use has been ``expressly reserved by their rightholders in an appropriate manner, such as machine-readable means in the case of content made publicly available online'' (Art.~4(3)). The AI Act requires providers of general-purpose AI models to identify and comply with these reservations \citep[Art.~53(1)(c)]{aiact2024}. No harmonised standard defines an appropriate opt-out \citep{lucchi2025,euipo2025}, while courts disagree on whether statements in natural language suffice \citep{olg2025,quintais2025}. The five countries implemented Article~4(3) with different wording, from a requirement of machine-readable form with metadata in Poland to no requirement of form in Slovakia.

\subsection*{National implementation and case law}
Article~7(1) of the Directive protects Articles~3, 5 and 6 against contractual override but not Article~4, so the terms of a website may also exclude the general exception. The five countries implemented the Directive with different wording (Table~\ref{tab:national}), a divergence that was noted soon after the deadline \citep{MargoniKretschmer2022}. German law makes a reservation for online works effective only in machine-readable form. The Polish provision, added in 2024 \citep{PLAmend2024}, requires a machine-readable format together with metadata. The Czech and Hungarian texts require machine-readable means, the Hungarian one as an advance objection. The Slovak text states no requirement of form, although it has to be read in conformity with the Directive.

\begin{table}[htbp]
\centering
\setlength{\tabcolsep}{3pt}
\caption{National provisions on reserving text and data mining}
\label{tab:national}
\begin{tabularx}{\textwidth}{@{}l >{\raggedright\arraybackslash}p{2.9cm} X@{}}
\toprule
 & Provision & Condition for content available online \\
\midrule
DE & \S~44b(3) UrhG \citep{law_de_urhg} & ``nur dann wirksam, wenn er in maschinenlesbarer Form erfolgt'' (effective only if made in machine-readable form) \\
PL & art.~26$^3$ ust.~2 pr.\ aut. \citep{law_pl_praut,law_pl_opendata} & express reservation ``w formacie przeznaczonym do odczytu maszynowego'' (in a machine-readable format) as defined in the 2021 open data act ``wraz z metadanymi'' (with metadata) \\
CZ & \S~39c(2) z.~121/2000~Sb. \citep{law_cz_az} & ``v\'yslovn\v{e} vyhradil vhodn\'ym zp\r{u}sobem'' (expressly reserved in an appropriate manner) and online ``strojov\v{e} \v{c}iteln\'ymi prost\v{r}edky'' (machine-readable means) \\
HU & \S~35/A(1)(b) Szjt. \citep{law_hu_szjt} & advance objection in an appropriate manner ``g\'eppel olvashat\'o form\'aban'' (machine-readable form) \\
SK & \S~51c(1) z.~185/2015~Z.~z. \citep{law_sk_az} & ``ak tak\'eto pou\v{z}itie nie je v\'yslovne vyhraden\'e'' (if not expressly reserved), no requirement of form \\
\bottomrule
\end{tabularx}
\end{table}

The Code of Practice states that adherence to it does not constitute compliance with Union law on copyright (Recital~(a)) and keeps reservations expressed by other appropriate means effective \citep[Measure~1.3(2)]{cop2025}. In December 2025 the Commission opened a consultation on reservation protocols \citep{commission2025consultation}. No such list had been published by October 2026. In \emph{Like Company v Google Ireland} (C-250/25) a Hungarian court asked the Court of Justice whether the training of a language model involves reproduction and whether Article~4 applies to it \citep{C250_25_reference}. In \emph{Kneschke v LAION} the Regional Court of Hamburg remarked in an obiter dictum that disclaimers in natural language may be machine-readable \citep{LGHamburg2024LAION}, a view that \citet[p.~8]{quintais2025} calls ``contrary to Recital~18''. On appeal, the Higher Regional Court of Hamburg called the question ``umstritten'' (disputed), described the requirement as ``technikoffen'' (open to any technology) and listed robots.txt and the TDM Reservation Protocol among the forms under discussion \citep[p.~19]{olg2025}. It attributed a reservation by an image agency to the photographer, with the agency acting as a licensee within a contract \citep[p.~17]{olg2025}. The Federal Court of Justice scheduled the hearing of the revision for 3~September 2026 and had not decided by October 2026 \citep{BGH2026LAIONpm}. In a case on a news monitoring service, the District Court of Amsterdam found a robots.txt that excluded only certain AI crawlers insufficient \citep[para.~4.33]{rbams2024}. The Budapest Court of Appeal treated robots.txt as a machine-readable objection in a case on the indexing of a website by a search engine \citep{FIT2024robots,mezei2025}.

\bibliographystyle{apacite}
\bibliography{references}

\end{document}